\documentclass[twocolumn,trackchanges]{aastex701}

\usepackage{comment}
\usepackage{orcidlink}

\begin{document}

\title{Cloud Scale Star Formation and Gas Scaling Relations in the Milky Way}

\author[orcid=0009-0005-7533-9804,gname=Alphesunny, sname='Sarkar']{Alphesunny Sarkar} 
\affiliation{School of Astrophysics, Presidency University, 86/1 College Street, Kolkata 700073, India}
\email[show]{alphesunny2000@gmail.com}

\author[orcid=0000-0003-0295-6586,gname=Tapas, sname='Baug']{Tapas Baug} 
\affiliation{S. N. Bose National Centre for Basic Sciences, Block-JD, Sector-III, Salt Lake City, Kolkata 700106, India}
\email{tapasbaug@bose.res.in}

\author[orcid=0009-0003-6633-525X, gname=Ariful, sname='Hoque']{Ariful Hoque}
\affiliation{S. N. Bose National Centre for Basic Sciences, Block-JD, Sector-III, Salt Lake City, Kolkata 700106, India}
\email[]{arifulh882@gmail.com} 

\author[orcid=0000-0002-3236-2853,gname=Suchetana, sname='Chatterjee']{Suchetana Chatterjee} 
\affiliation{School of Astrophysics, Presidency University, 86/1 College Street, Kolkata 700073, India}
\email{suchetana.astro@presiuniv.ac.in}

\author[orcid=0000-0003-4531-0945, gname=Chayan, sname='Mondal']{Chayan Mondal} 
\affiliation{S. N. Bose National Centre for Basic Sciences, Block-JD, Sector-III, Salt Lake City, Kolkata 700106, India}
\email{cmondal@bose.res.in}

\begin{abstract}
We investigate cloud-scale star formation in the Milky Way using a sample of 45 molecular clouds (sizes of $5-240$ pc) in the inner Galactic plane, spanning heliocentric distances of $1.1-14.4$ kpc. Masses of these clouds are derived from $^{12}$CO and $^{13}$CO emission, while stellar masses are estimated using the young stellar object (YSO) population. The studied molecular clouds have masses ranging from $\sim10^{3}$ to $2.3\times10^{6}$ $\rm M_\odot$, with star formation efficiencies (SFE) up to 0.33. We find a tight, nearly linear scaling of the star formation rate (SFR) with the cloud mass, indicating that more massive clouds form proportionally more stars. The SFE, however, shows a declining trend with cloud mass. The relation between the star formation rate surface density ($\Sigma_{\rm SFR}$) and gas surface density ($\Sigma_{\rm gas}$) exhibits substantial cloud-to-cloud scatter, indicating that the canonical Kennicutt--Schmidt law is not strongly recovered at the scale of individual molecular clouds. Incorporating the cloud free-fall time ($\rm t_{ff}$) into the star formation scaling relation highlights its important role in regulating star formation, although the observed relations suggest that the star formation efficiency per free-fall time is not universal. In particular, the SFE decreases with increasing gas mass available per free-fall time. We discuss the implications of our results in the context of recent theoretical models of molecular cloud evolution and star formation.
\end{abstract}


\section{Introduction} 

Understanding the star-formation at different spatial scales and at different epochs in the Universe, stands as one of the most crucial questions pertaining to studies of structure formation \citep{McKee2007Review, Kennicutt2012Review}. Quantifying star formation requires accurate estimates of both the stellar population and the molecular gas reservoir. These quantities form the basis for determining the star formation rate (SFR) and star formation efficiency (SFE). Different tracers are used to estimate SFR and SFE in Galactic to extra-galactic scales. The most widely known tracers are UV and IR emission in galaxies \citep{Buat1999SFR_UVIR,Hirashita2003SFR_UVIR}, H$\alpha$ emission \citep{Kennicutt2012Review, Altenburg2007Halpha}, and extragalactic background light \citep[EBL;][]{Koushan2021EBL, Driver2016EBL, Andrews2018EBL}. However for the Milky Way and other nearby spatially resolved galaxies, counting Young Stellar Objects (YSOs) provide a more direct and reliable indicator of star-formation \citep{Kennicutt2012Review}, since the YSO-counting method directly probes stellar mass rather than relying on secondary emissions. 

The dependence of SFR on gas mass was first proposed by \cite{Schmidt1959LAW}, who argued that the SFR volume density is proportional to the square of the gas density in the Milky Way galaxy. \cite{Kennicutt1998LAW} expanded this idea to global scales using a sample of 61 spiral galaxies and 36 starburst galaxies. Since observations of external galaxies generally measure surface densities integrated along the line of sight, the relation is commonly expressed in terms of SFR surface density ($\Sigma_\mathrm{SFR}$) and surface density of gas ($\Sigma_\mathrm{gas}$). This formulation also reduces the influence of the common dependence of gas mass and star formation rate on the overall size of the system. A tight power law relation $\Sigma_\mathrm{SFR}\propto \Sigma_\mathrm{gas}^{N}$, with a power law exponent of $N=1.4\pm0.15$ was observed \citep{Kennicutt1998LAW}. This is known as the ``Kennicutt-Schmidt'' law, where $\Sigma_\mathrm{gas}$ was calculated by including both atomic (H{\sc i}) and molecular ($\mathrm{H_2}$) hydrogen. Subsequently \cite{Bigiel2008LAW} and \cite{Leroy2013LAW} examined this scaling relation in nearby resolved spiral and disk galaxies at sub-kpc to kpc resolutions and showed that when molecular gas ($\mathrm{H_2}$) alone is considered, the relationship becomes nearly linear ($N\approx1$). \cite{Krumholz2009LAW} developed a theoretical framework considering star formation to be regulated by molecular gas content. In high surface density systems such as starbursts, their relation reduces to $\Sigma_\mathrm{SFR}\propto \Sigma_\mathrm{gas}^{1.33}$ which is consistent with the Kennicutt slope within uncertainties. In molecular gas dominated regimes they predict almost a linear relation.

A wide range of power-law exponents has been reported for extragalactic systems, from individual galaxies to galaxy groups, spanning $N=0.8 - 3.3$ \citep{Boissier2003LAW,Heyer2004LAW,Komugi2005LAW,Blanc2009LAW, Roychowdhury2015Scaling, Wilson2019Scaling, Kennicutt2021Scaling}. In contrast, several works examined this scaling relation in the Milky Way from cloud scales to clump scales, representing the parent star-forming structures and their dense substructures respectively \citep{Wu2005LAW, Evans2009LAW, Heiderman2010LAW, Lada2010SFR, Vutisalchavakul2016SFR, Lee2016LAW, Pokhrel2021LAW, Das2021NANComplex, Wells2022LAW, Elia2022LAW, Rawat2025LAW}, but only a few studies explicitly quantified the power-law slope. For example, on clump scales \cite{Wu2005LAW} reported a linear relation ($N\approx 1)$ while \cite{Rawat2025LAW} obtained an exponent of $N=1.46$. On cloud scales \cite{Pokhrel2021LAW} found a steeper slope of $N=2$. 

Although the `Kennicutt-Schmidt' scaling relation has been widely tested across clump, cloud to extragalactic scales, \cite{Krumholz2012Freefall} argued that  incorporating the free-fall time ($\mathrm{t_{ff}}$) leads to a more physically motivated star formation relation. They proposed that $\mathrm{\Sigma_{SFR}}$ correlates more tightly with $\mathrm{\Sigma_{gas}/t_{ff}}$, and they expressed this volumetric star formation relation as $\mathrm{\Sigma_{SFR}}=\mathrm{\epsilon_{ff}\Sigma_{gas}/t_{ff}}$ where $\mathrm{\epsilon_{ff}}$ is the efficiency per free-fall time. The basis of this volumetric star-formation law is based on the analysis of \citet{k&M05}, where they argue that supersonic turbulence prevailing in molecular clouds exhibits a self-similar turbulent velocity dispersion relation for a given length scale. The volumetric star-formation relation has been subsequently examined at both clump \citep{Das2021NANComplex,Rawat2025LAW} and cloud \citep{Vutisalchavakul2016SFR,Pokhrel2021LAW} scales yielding sub-linear to linear scaling with $\mathrm{\epsilon_{ff}\sim 1-3\% }$. 

The SFE is commonly defined as the ratio of the SFR to the available molecular gas mass \citep{Kennicutt2012Review}. On galaxy-wide scales this quantity is observed to vary weakly among different galaxy populations \citep{GaoANDSolomon2004SFRSFE}. However, SFE is often defined as the fraction of cloud or clump mass converted into stars, on smaller Galactic scales \citep{Myers1986SFEdef}. Studies have shown that the SFE decreases with increasing cloud or clump mass \citep{Wells2022LAW}. Typical SFE values range from $8-20\%$ for clumps \citep{Wells2022LAW, Rawat2025LAW} and $1-6\%$ for molecular clouds \citep{Evans2009LAW, Heiderman2010LAW, Lada2010SFR, Rawat2025LAW} with masses up to $10^5$ $\mathrm{M_{\odot}}$. Most of these cloud-scale studies have focused on nearby, relatively low-mass star-forming regions. Since the physical conditions governing star formation vary throughout the Galaxy, both the SFR and SFE are expected to exhibit significant spatial variations \citep{Kennicutt2012}.

Keeping these issues at hand we perform a wider study of star-formation properties in Galactic clouds. Our data spans a broader region of the Galaxy, i.e., in longitudes $l\sim10-50^{\circ}$ and $300-350^{\circ}$ over distances $d\sim1.1-14.4$ kpc. Previous cloud-scale studies were largely restricted to nearby molecular clouds, typically within a few hundred parsecs of the Sun, with masses of order $10^3-10^5$ $\mathrm{M_{\odot}}$ \citep{Evans2009LAW,Heiderman2010LAW,Lada2010SFR}. Later studies of more distant Galactic plane clouds expanded the explored mass range to $3.3\times 10^5$ $\mathrm{M_{\odot}}$ \citep{Vutisalchavakul2016SFR, Retes-Romero2017LAW}. In comparison, the molecular clouds analyzed in this work extend up to $2.3\times 10^6$ $\mathrm{M_{\odot}}$, which is an order of magnitude higher. Finally, we examine the volumetric star-formation relations for our sample and study their physical connections with the Krumholz and other alternative models. The paper is organized as follows: In the next section, we describe our datasets and methodology. We describe and discuss our results in Sections \ref{sec: Results} and \ref{sec: Discussions} respectively.

\section{Data Sets and Methodology} \label{sec: Data}
We now discuss our data sets for estimating the SFR and SFE in different star-forming clouds in the inner Galactic plane. This study is primarily based on the estimation of stellar mass with the identification of YSOs. For that, we employ a number of infrared telescopes and surveys namely i) Spitzer, ii) WISE, iii) 2MASS, iv) Herschel and v) AKARI/APEX. Below we give a brief description of our infrared datasets. 

\subsection{Photometric data}
We utilize the data from the Galactic Legacy Infrared Midplane Survey Extraordinaire \citep[GLIMPSE; $l=10^{\circ}-65^{\circ}$ and $295^{\circ}-350^{\circ}$, $ b=\pm 1^{\circ}$;][]{Benjamin2003GLIMPSE, Churchwell2009Spitzer}. The four Infrared Array Camera \citep [IRAC;][] {Fazio2004IRAC} bands provide the data at wavelengths 3.6, 4.5, 5.8, and 8.0 $\mu \mathrm{m}$ with resolution $1\farcs7$, $1\farcs7$, $1\farcs8$, $2\farcs0$ respectively. The 24 $\mu \mathrm{m}$ data are taken from MIPS Galaxy survey \citep [MIPSGAL; covering: $ l<69^{\circ}$ and $ l>292^{\circ}$ with $|b|\leq 1^{\circ}$;][] {Carey2009MIPSGAL, Rieke2004MIPS}. It has a resolution of $6\farcs0$.

The Wide-ﬁeld Infrared Survey Explorer (WISE) uses the four infrared bands 3.4 $\mu \mathrm{m}$ (W1), 4.5 $\mu \mathrm{m}$ (W2), 12 $\mu \mathrm{m}$ (W3), 22 $\mu \mathrm{m}$ (W4) to survey the entire sky \citep{Wright2010WISE}. The angular resolution of W1, W2, W3 and W4 are $6\farcs1$, $6\farcs4$, $6\farcs5$ and $12\farcs0$ respectively. The data provided for W1, W2, W3 and W4 are in magnitudes, and we convert them into flux densities using the zero magnitude flux values from \citet{Jarrett2011SpitzerWise}.

The Two Micron All Sky Survey (2MASS) \citep{Skrutskie2006twoMASS,Cutri2003twoMASSCatalog} uses three bands $J$ (1.235 $\mu \mathrm{m}$), $H$ (1.662 $\mu \mathrm{m}$) and $K_S$ (2.159 $\mu \mathrm{m}$). It has $2\farcs5$ resolution in each of the three bands.

The Herschel Space Observatory \citep{Pilbratt2010Herschel} has two main instruments, Photodetector Array Camera and Spectrometer (PACS) and Spectral and Photometric Imaging REceiver (SPIRE) to cover far infrared to submillimeter bands. PACS covers 70 and 160 $\mu \mathrm{m}$ bands  with resolutions $6\farcs0$ and $13\farcs0$ respectively and SPIRE covers 250, 350, 500 $\mu \mathrm{m}$ bands with resolutions $18\farcs0$, $25\farcs0$, $36\farcs0$, respectively. We took data from the Herschel infrared Galactic Plane Survey \citep[Hi-GAL;][]{Molinari2016Hi-GAL} for this study. Hi-GAL covers $-70^{\circ}\leq l\leq 68^{\circ}$ and $|b|\leq 1^{\circ}$ in the inner Milky Way.

The complementary data are obtained from Infrared Astronomical Satellite mission \citep[AKARI;][]{Murakami2007AKARI}. The Infrared Camera (IRC) provides all sky survey at 9 $\mu \mathrm{m}$ ($5\farcs7$) and 18 $\mu \mathrm{m}$ ($5\farcs5$) wavelengths. We utilize the  All-Sky Survey Point Source Catalog data in our study \citep{Ishihara2010AKARI}. The higher wavelength data at 870 $\mu \mathrm{m}$ is taken from the APEX Telescope Large Area Survey of the Galaxy \citep[ATLASGAL;][]{Schuller2009ATLASGAL}. The survey was conducted using the Large APEX Bolometer Camera (LABOCA) with a resolution of $19\farcs2$. All the above data except ATLASGAL are taken from the NASA/IPAC Infrared Science Archive (IRSA)\footnote{\url{https://irsa.ipac.caltech.edu/applications/Gator/}}, and ATLASGAL data are taken from the ATLASGAL database\footnote{\url{https://atlasgal.mpifr-bonn.mpg.de/cgi-bin/ATLASGAL_DATASETS.cgi}}.

\subsection{CO line Data}
To estimate the cloud masses, we used $^{12}$CO ($J=1-0$) and $^{13}$CO ($J=1-0$) spectral line data cubes obtained from two surveys: The FOREST Unbiased Galactic plane Imaging survey with the Nobeyama 45-m telescope \citep[FUGIN;][] {Umemoto2017FUGIN} and The Three-mm Ultimate Mopra Milky Way Survey \citep[ThrUMMS;][]{Barnes2015ThrUMMS}.

The Galactic longitude range $l\sim10-50^{\circ}$ was covered by the FUGIN survey. This survey has angular resolution of $20\farcs0$ and velocity resolution of 1.3 $\mathrm{km\, s^{-1}}$. The average rms noise levels in the main beam temperature for the above velocity resolution are $\sim 1.47$ K for $^{12}$CO and $\sim 0.69$ K for $^{13}$CO with $8\farcs5$ pixels. The Galactic longitude range $l\sim300-360^{\circ}$ was covered by the ThrUMMS survey, which has the angular and velocity resolutions of $72\farcs0$ and 0.3 $\mathrm{km\,s^{-1}}$ respectively.

\begin{figure}[t]
    \centering
    \includegraphics[width=1.0\columnwidth, height=0.3\textheight]{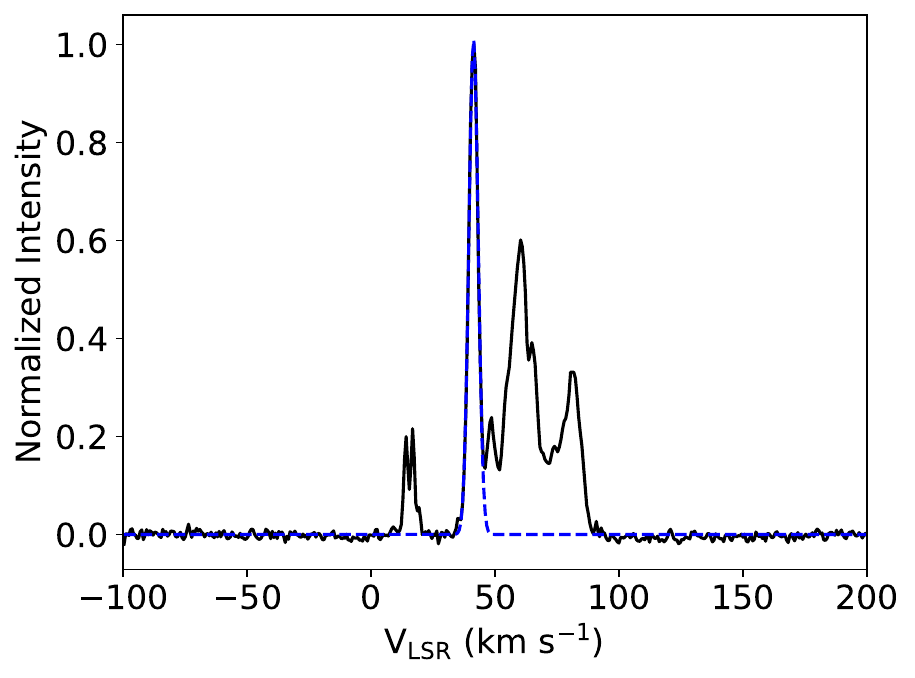} 
    \caption{$^{13}$CO spectrum of the cloud G038.925-00.355. The black line represents the observed emission and the blue dashed line represents the best-fit Gaussian profile.}
    \label{fig: spectrum}
\end{figure}
  
\begin{figure*}[!t]
    \centering
    \includegraphics[width=1.3\columnwidth, height=1.3\columnwidth]{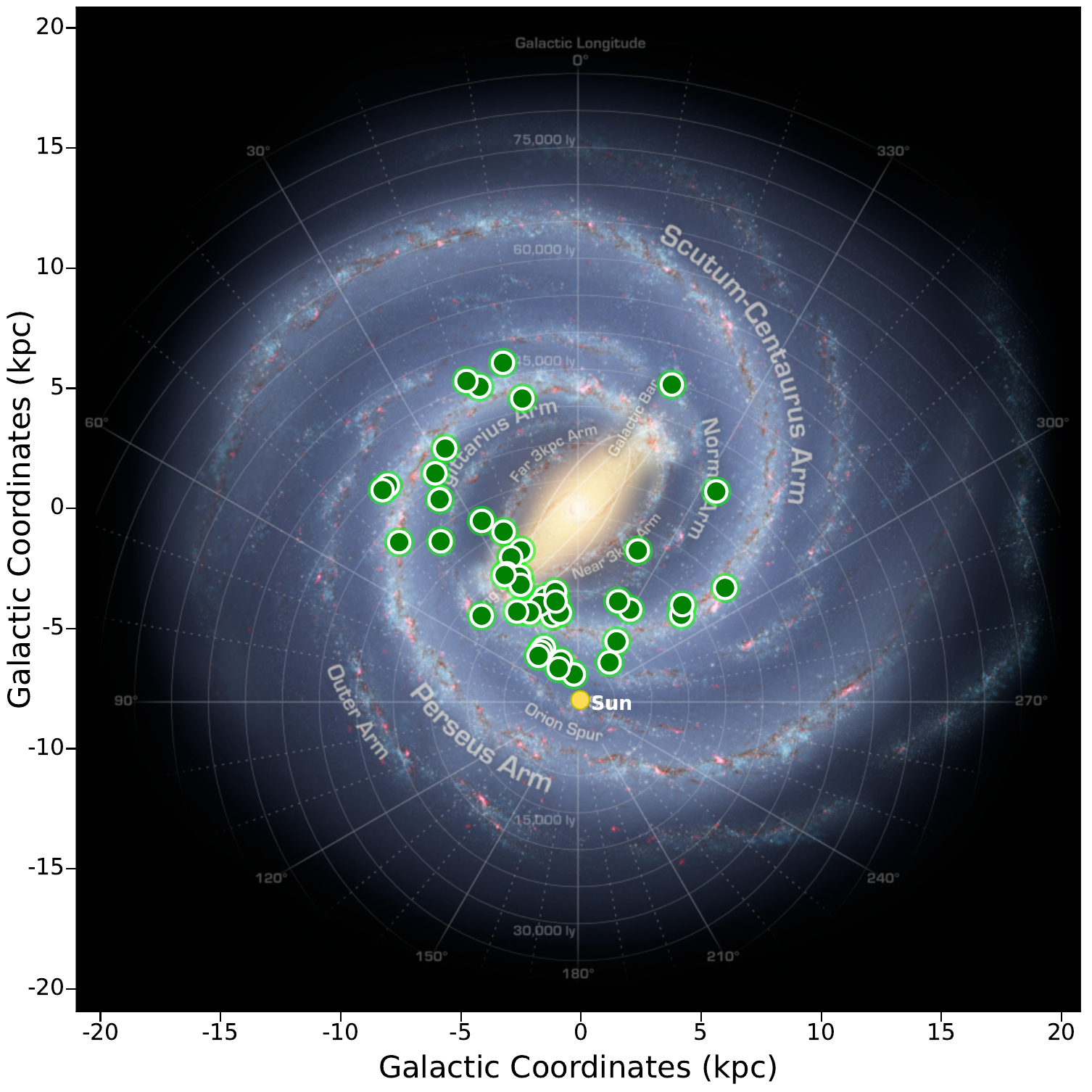} 
    \caption{The Milky Way image: overlaid on it the 45 star-forming clouds (blue filled circles) with the yellow filled circle representing the Sun. Background image credit: NASA/JPL-Caltech/R. Hurt (SSC/Caltech).}
    \label{fig:MilkyWay}
\end{figure*}

\subsection{Identification of the Clouds}
The potential clouds are selected primarily from the catalog of \cite{Anderrson2014Catalog}. These authors report approximately 8000 H{\sc ii} region candidates in the Galactic plane along with their Galactic coordinates, Galactocentric distances (${R_{\rm gal}}$), local standard of rest velocities ($\mathrm{V_{LSR}}$) and heliocentric distances. We first select sources for which coordinates and kinematic parameters are available. The sample is subsequently restricted to regions lying within the coverage of the FUGIN and ThrUMMS CO surveys, since $^{12}$CO and $^{13}$CO data products are required to estimate the cloud parameters. To avoid bias toward any particular Galactic longitude range, the selected sources were distributed across the full longitudinal extent of the survey area. In addition, overlapping or confused regions were excluded to obtain reliable measurements of cloud properties and the associated YSO populations. These criteria reduced the initial catalog to a final sample of 24 clouds.

In order to increase the sample size we followed the spectral identification method using our CO data. We extracted the spectrum (Figure~\ref{fig: spectrum}) from the data cube that gives us several peak $\mathrm{V_{LSR}}$ corresponding to several clouds. For each peak $\mathrm{V_{LSR}}$, we visually define a central $l$, $b$ by inspecting the cloud. In cases where multiple velocity components are present, we select the component whose spatial morphology is consistent with the YSO distribution. Using this procedure, a total of 21 clouds are identified. The Galactocentric and heliocentric distances of these clouds are taken from the literature \citep{Cabral2021SEDIGISM, Bowers2015Bolocam, Rigby2019CHIMPS}. We also confirmed the distance measurements (Table \ref{table:sources}) of these clouds using the ``Kinematic Distance Calculation Tool"\footnote{\url{https://www.treywenger.com/kd/index.php}} \citep{TreyWenger2018KDE}, which employs Monte Carlo techniques to determine kinematic distances considering the galaxy rotation curve \citep{Reid2014KDE}. The details of 45 clouds (combining the two selection procedures) are listed in Table~\ref{table:sources}. In Figure~\ref{fig:MilkyWay}, we overlaid the positions of the clouds onto an artistic image of the Milky Way to show their distribution throughout the Galaxy.

To define the boundary of each cloud we adopted the following procedure. The $^{13}$CO spectrum for each cloud was extracted and fitted with a Gaussian profile (Figure~\ref{fig: spectrum}) to determine the $\rm{V_{LSR}}$ and the Gaussian width $\sigma$. The H$_2$ column density map was then generated by integrating the emission within the velocity range $\rm{V_{LSR}}\pm3\sigma$. Details of the H$_2$ column density calculation are presented in Section~\ref{sec:column density}. A cloud boundary was defined using the contour corresponding to three times the local background column density (Figure~\ref{fig: Column density map}), where the background level is estimated from a nearby emission free region via visual inspection. The resulting cloud boundary thresholds range from $3.15\times10^{20}$ to $2.79\times10^{21}$ cm$^{-2}$ across our sample.

\begin{deluxetable*}{lccccccc}[htb!]
\tablecaption{Details of our Cloud Sample}
\tablewidth{0pt}
\tabletypesize{\footnotesize}
\tablehead{
\colhead{Cloud} & \colhead{$l$} & \colhead{$b$} & \colhead{V$_{\rm LSR}$} & \colhead{V$_{\rm range}$} & \colhead{$d$} & \colhead{$R_{\rm gal}$} \\
 & \colhead{(deg)} & \colhead{(deg)} & \colhead{(km s$^{-1}$)} & \colhead{(km s$^{-1}$)} & \colhead{(kpc)} & \colhead{(kpc)}
}
\startdata
G010.560-00.080$^*$ & 10.744 & -0.189 & 29.7 & [16.7, 42.0] & $12.8\pm0.6$ & $4.8\pm 0.5$ \\
G012.774+00.330 & 12.776 & 0.551 & 18.0 & [14.0, 23.2] & $14.4\pm0.8$ & $6.4\pm 0.6$ \\
G012.870-00.255$^*$ & 13.057 & -0.245 & 36.8 & [25.8, 42.3] & $3.7\pm0.5$ & $4.8\pm0.5$ \\
G013.272+00.066$^*$ & 12.966 & -0.082 & 53.1 & [46.0, 59.6] & $4.6\pm0.5$ & $3.9\pm0.4$ \\
G013.913+00.239$^*$ & 13.956 & 0.247 & 47.2 & [40.8, 53.1] & $4.2\pm0.4$ & $4.3\pm0.4$ \\
G014.331-00.640 & 13.999 & -0.475 & 22.0 & [14.0, 26.4] & $1.1\pm0.1$ & $7.4\pm0.5$ \\
G016.285-00.171 & 16.837 & -0.281 & 47.6 & [37.4, 56.3] & $4.1\pm0.4$ & $4.7\pm0.4$ \\
G017.638+00.157 & 17.707 & 0.214 & 22.2 & [15.4, 31.0] & $13.7\pm0.6$ & $6.2\pm0.5$ \\
G018.371-00.382 & 18.116 & -0.389 & 44.3 & [34.2, 60.2] & $3.7\pm0.4$ & $5.2\pm0.4$ \\
G018.824-00.467 & 18.925 & -0.398 & 63.2 & [56.3, 75.2] & $4.5\pm0.4$ & $4.5\pm0.3$ \\
G019.480+00.155 & 19.501 & 0.065 & 22.2 & [14.0, 36.8] & $14.1\pm0.6$ & $6.7\pm0.4$ \\
G021.596-00.161 & 21.495 & -0.105 & 120.0 & [115.5, 125.9] & $6.7\pm0.1$ & $3.4\pm0.2$ \\
G023.253-00.240 & 22.976 & -0.316 & 63.4 & [48.5, 79.7] & $4.3\pm0.4$ & $4.9\pm0.3$ \\
G024.426+00.221 & 24.355 & 0.255 & 116.7 & [98.6, 127.8] & $7.7\pm1.0$ & $3.7\pm0.2$ \\
G025.796+00.239 & 25.717 & 0.243 & 109.9 & [99.9, 120.7] & $6.6\pm0.1$ & $3.9\pm0.3$ \\
G026.272-00.062$^*$ & 26.223 & 0.033 & 99.9 & [88.9, 114.2] & $5.7\pm0.6$ & $4.0\pm0.3$ \\
G026.597-00.024 & 26.505 & -0.045 & 22.9 & [18.0, 33.6] & $1.8\pm0.6$ & $6.9\pm0.4$ \\
G027.000+00.190$^*$ & 27.227 & 0.090 & 94.0 & [85.0, 101.9] & $5.4\pm0.7$ & $4.2\pm0.3$ \\
G027.996-00.412$^*$ & 28.110 & -0.385 & 44.0 & [41.4, 50.5] & $11.9\pm0.5$ & $5.9\pm0.3$ \\
G028.652+00.072$^*$ & 28.652 & 0.023 & 107.7 & [96.6, 114.9] & $8.5\pm0.6$ & $4.1\pm0.2$ \\
G029.832-00.100 & 29.910 & -0.052 & 99.1 & [88.2, 112.2] & $6.1\pm0.8$ & $4.4\pm0.3$ \\
G030.000-00.335$^*$ & 29.877 & -0.407 & 70.6 & [64.1, 76.5] & $4.2\pm0.5$ & $5.1\pm0.3$ \\
G031.200+00.119$^*$ & 31.039 & 0.172 & 107.0 & [94.0, 116.8] & $6.1\pm0.6$ & $4.5\pm0.1$ \\
G032.473+00.204 & 32.472 & 0.274 & 49.5 & [39.4, 60.2] & $11.2\pm0.4$ & $6.1\pm0.3$ \\
G034.404+00.227 & 34.100 & 0.096 & 57.4 & [48.5, 66.0] & $1.6\pm0.1$ & $7.3\pm0.3$ \\
G034.758-00.525$^*$ & 34.778 & -0.691 & 42.0 & [32.3, 57.6] & $2.6\pm0.4$ & $6.4\pm0.3$ \\
G035.063+00.330 & 34.896 & 0.312 & 53.1 & [40.0, 64.8] & $10.2\pm0.4$ & $5.9\pm0.3$ \\
G035.480+00.135$^*$ & 35.402 & 0.168 & 77.1 & [66.7, 90.1] & $4.5\pm0.7$ & $5.2\pm0.3$ \\
G038.925-00.355$^*$ & 38.906 & -0.458 & 40.7 & [35.5, 46.6] & $2.6\pm0.4$ & $6.6\pm0.3$ \\
G041.184-00.210$^*$ & 41.203 & -0.281 & 59.6 & [51.1, 68.7] & $8.8\pm0.6$ & $6.0\pm0.3$ \\
G041.740+00.095 & 41.754 & 0.056 & 13.2 & [8.9, 25.1] & $12.0\pm0.7$ & $8.0\pm0.3$ \\
G043.149+00.028 & 43.149 & -0.051 & 11.8 & [-0.9, 25.1] & $12.0\pm0.7$ & $8.2\pm0.3$ \\
G043.225-00.312$^*$ & 43.208 & -0.253 & 40.7 & [34.9, 48.5] & $2.5\pm0.5$ & $6.6\pm0.3$ \\
G048.599+00.044 & 48.781 & 0.075 & 17.7 & [9.5, 23.9] & $10.0\pm0.8$ & $7.7\pm0.3$ \\
G049.428-00.464 & 49.368 & -0.241 & 57.5 & [44.6, 75.2] & $5.4\pm0.3$ & $6.5\pm0.2$ \\
G307.610-00.382$^*$ & 307.631 & -0.462 & -35.0 & [-41.7, -28.9] & $7.6\pm0.7$ & $7.0\pm0.2$ \\
G310.143+00.758 & 310.034 & 0.427 & -55.3 & [-58.9, -52.0] & $5.5\pm0.1$ & $6.5\pm0.2$ \\
G312.598+00.048 & 312.809 & 0.220 & -62.7 & [-70.2, -53.7] & $5.8\pm0.1$ & $6.3\pm0.2$ \\
G321.890-00.009$^*$ & 321.954 & -0.043 & -31.2 & [-36.3, -27.2] & $2.0\pm0.4$ & $6.9\pm0.3$ \\
G327.129-00.257$^*$ & 326.768 & -0.289 & -60.0 & [-69.2, -52.3] & $10.4\pm0.5$ & $5.6\pm0.3$ \\
G328.221-00.531$^*$ & 328.136 & -0.531 & -43.8 & [-52.3, -34.7] & $2.9\pm0.4$ & $6.1\pm0.3$ \\
G331.123-00.530 & 331.025 & -0.410 & -66.5 & [-74.8, -57.9] & $4.3\pm0.4$ & $5.2\pm0.3$ \\
G338.911+00.615 & 338.849 & 0.593 & -62.1 & [-72.6, -51.7] & $4.4\pm0.4$ & $4.7\pm0.4$ \\
G339.010-00.138$^*$ & 338.818 & -0.262 & -118.0 & [-128.0, -108.5] & $6.7\pm0.5$ & $3.1\pm0.2$ \\
G343.873-00.078$^*$ & 343.693 & -0.375 & -23.6 & [-33.2, -16.8] & $13.7\pm0.5$ & $6.1\pm0.5$ \\
\enddata
\tablecomments{Column (1): Cloud name. Columns (2)–(3): Galactic coordinates representing the centre of the region. Column (4): Local standard of rest velocity. Column(5): Velocity range used to construct column density map. Column (6): Kinematic distance with uncertainty. Column (7): Galactocentric distance with uncertainty.\\
$^*$ Clouds are selected from spectral identification.}
\label{table:sources}
\end{deluxetable*}

\subsection{Column Density Map and Cloud Mass}\label{sec:column density}

\begin{figure}[t]
    \centering
    \includegraphics[width=1.0\columnwidth, height=0.35\textheight]{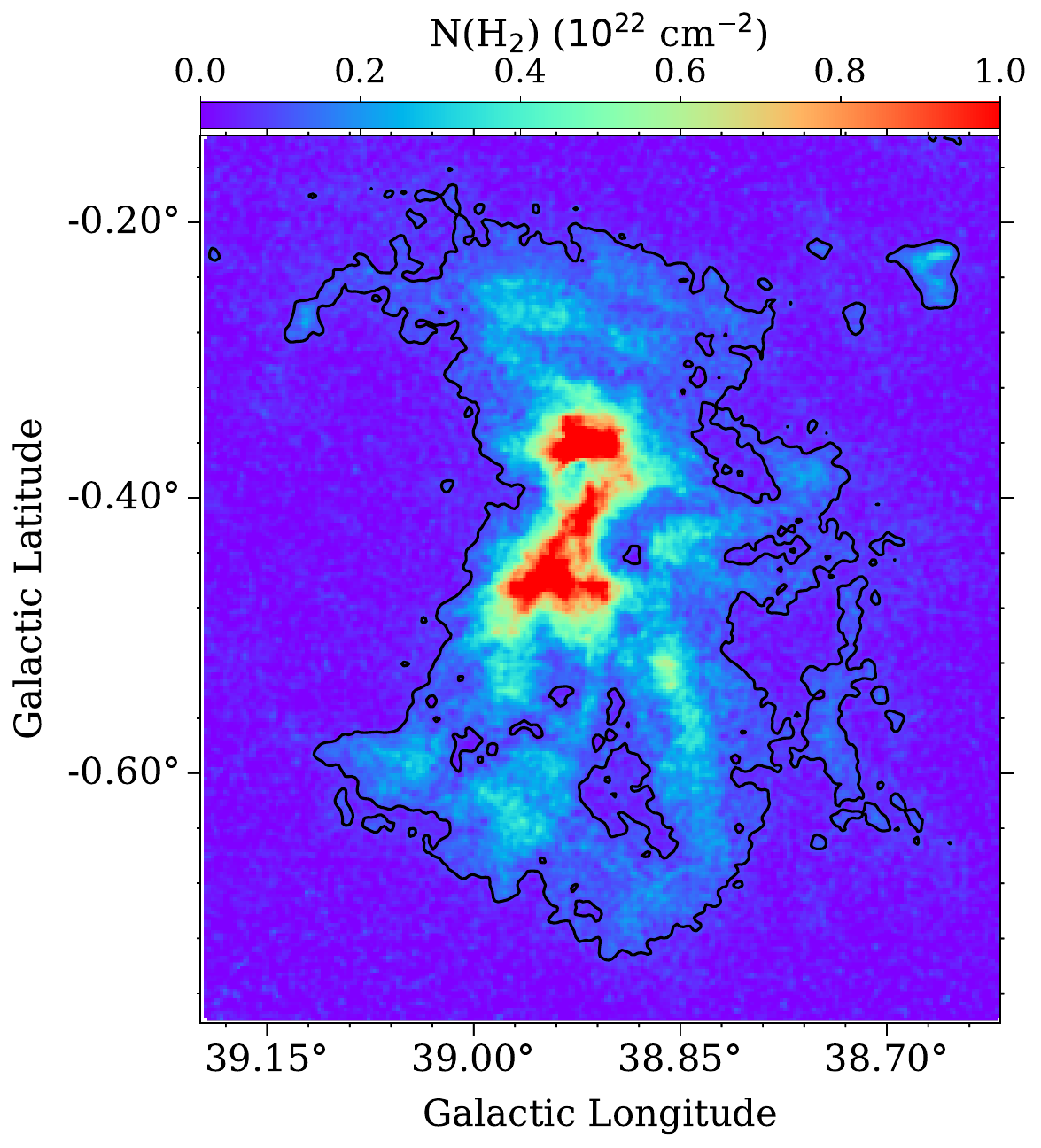} 
    \caption{H$_2$ column density map of the cloud G038.925-00.355. The black contour marks the cloud boundary defined by a threshold of $8.43\times10^{20}$ cm$^{-2}$, corresponding to three times the background column density.}
    \label{fig: Column density map}
\end{figure}

The column density map (Figure~\ref{fig: Column density map}) for each cloud was generated using $^{12}$CO and $^{13}$CO spectral cubes considering local thermodynamic equilibrium (LTE) (\citealt{Wilson2009Book}). The excitation temperature ($T_\mathrm{ex}$) was obtained from the $^{12}$CO cube assuming that the transition of $^{12}$CO is optically thick.

\begin{equation}
    T_\mathrm{ex}=5.5\Big/ \ln \left[1+ \frac{5.5}{T_\mathrm{mb}(^{12}\mathrm{CO\ peak)}+0.82} \right]\;\; [\mathrm{K}],
\end{equation}
where $T_{\mathrm{mb}}(^{12}\mathrm{CO\ peak})$ is the $^{12}$CO peak intensity. 

This $T_\mathrm{ex}$ and the $^{13}$CO brightness temperature ($T_{\mathrm{mb}}(\nu)$) are further utilized to estimate the optical depth of $^{13}$CO emission ($\tau_{13}(\nu$)) at each pixel and velocity channel, under the assumption that the two different isotopic species share the same excitation temperature \citep{Kohno2021CO}:

\begin{equation}
\tau_{13}(\nu) = -\ln \left[\, 1 - \frac{T_{\mathrm{mb}}(\nu)}{5.3}
\left( \frac{1}{\exp\left(\frac{5.3}{T_{\mathrm{ex}}}\right) - 1} - 0.16 \right)^{-1} \right]
\end{equation} 

The $^{13}$CO column density [$N(^{13}$CO)] at each pixel is obtained by integrating the emission over the defined velocity range (see \citealt{Kohno2021CO} for more details) :

\begin{equation}
N(^{13}\mathrm{CO}) = 2.4 \times 10^{14} \times
\sum \frac{T_{\mathrm{ex}}\, \tau_{13}(\nu)\, \Delta \nu}
{1 - \exp\!\left(-\frac{5.3}{T_{\mathrm{ex}}}\right)}
\;\; [\mathrm{cm}^{-2}].
\end{equation} 

Since the clouds in our sample span a wide range in Galactocentric radius, we account for the radial variation in molecular abundance and isotopic ratio by adopting a metallicity-dependent conversion factor $X(R_{\rm gal},Z)$ proposed by \citet{Patra2025_conversion_factor}. The hydrogen column density is therefore calculated as $N(\mathrm{H}_2)=X(R_{\rm gal},Z)\,N(^{13}\mathrm{CO})$, where

\begin{equation}
    X(R_{\rm gal},Z)=6000\left[ \frac{1}{Z}\right] \left(5.87\times \frac{R_{\rm gal}}{\rm kpc}+13.25\right),
\end{equation}

Here, the $R_{\rm gal}$ dependent term represents the Galactocentric variation of $^{12}\rm CO/^{13}CO$ isotopic abundance ratio following \citet{Jacob2020Isotopic_abundance}. The metallicity $Z$ of each cloud in our sample was estimated as follows. \citet{Delgado2022AbundanceGradient} investigated the chemical abundance gradients of several elements (He, C, N, O, Ne, S, Cl, and Ar) in Galactic H{\sc ii} regions, both with and without corrections for temperature fluctuations ($t^2$). We adopt their temperature-corrected oxygen gradient expressed as 12+log(O/H)=${9.22-0.059R_{\rm gal}}$, as it accounts for temperature inhomogeneities and provides abundances that are not affected by the abundance discrepancy problem. Since oxygen is the dominant metal content \citep{Osterbrock1989Book}, the resulting oxygen abundance is converted to metallicity using $Z=Z_{\odot}\frac{\mathrm{O/H}}{\mathrm{(O/H)_{\odot}}}$, adopting $Z_{\odot}=0.0143$ and 12+log(O/H)=8.50 for the solar neighborhood \citep{Patra2024Metallicity}.

Finally, the cloud masses are estimated using the following equation:

\begin{equation}
    M = \mu_\mathrm{H_2} \, m_{\mathrm{H}} \, A_{\mathrm{pix}} \sum N(\mathrm{H}_2),
\end{equation}
where the mean molecular weight of hydrogen ($\mu_\mathrm{H_2})$ is 2.8, $m_\mathrm{H}=1.67\times10^{-24}$g and $A_\mathrm{pix}$ is the area subtended by a single pixel. The estimated masses of the star forming clouds are listed in Table~\ref{tab:CloudMass}. 

\subsection{YSO Selection}

The identification and classification of YSOs in each cloud are performed using a three-phase infrared color-magnitude criterion developed and modified by \citet{Gutermuth2008Spitzer, Gutermuth2009Spitzer}. This analysis incorporates photometric measurements across eight bands \textit{Spitzer}/IRAC (3.6, 4.5, 5.8 and 8.0\,$\mu\mathrm{m}$), MIPS ($24\,\mu\mathrm{m}$) and 2MASS ($J$, $H$, $K_S$) to identify and classify YSOs. In phase one, all the sources detected in the four IRAC bands with photometric uncertainties $\sigma<0.2$ mag are considered. Contaminants such as active star forming galaxies, active galactic nuclei (AGN), knots of shock emission are removed by IRAC four band color criteria \citep{Gutermuth2009Spitzer}. The remaining sources are then classified into Class I, Class II and Class III/field stars.

In phase two, additional YSOs are identified from the previously classified field star and from the sources lacking detections at either 5.8 $\mu \mathrm{m}$ or 8.0 $\mu \mathrm{m}$ but possessing high quality 2MASS $H$ and $K_S$ photometry. Since the line of sight extinction reddens the observed color, this is corrected using the color excess ratio $\frac{E_{J-H}}{E_{H-K}}$ or $\frac{E_{3.6-4.5}}{E_{H-K}}$ (when J band photometry is unavailable) adopted from \cite{Flaherty2007Extinction}. The de-reddened K-[3.6] and [3.6]-[4.5] colors are then computed, allowing us to identify additional Class I and Class II sources.

In phase three, a refinement is performed using MIPS 24 $\mu \mathrm{m}$ photometry. All the sources with MIPS 24 $\mu \mathrm{m}$ uncertainties  $\sigma<0.2$ mag are considered. Among the previously classified sources some objects initially assigned as Class III are reclassified as `transition disk' which is an intermediate stage between Class II and Class III, characterized by substantial clearing of disk material. Sources that remain unidentified in phase one and phase two due to unreliable IRAC measurements are classified as deeply embedded protostars if they satisfy the criteria $[24]<7$ and $[X]-24>4.5$ mag, where $[X]$ denotes the longest wavelength IRAC band detection. Additional protostar candidates are also recovered from previously classified AGN and shock dominated sources. A fraction of initially flagged protostars are finally reclassified to the Class II category. Accordingly, our analysis is based on the final Class I and Class II YSO populations.

 \begin{figure}[t]
    \centering
    \includegraphics[width=1.0\columnwidth, height=0.3\textheight]{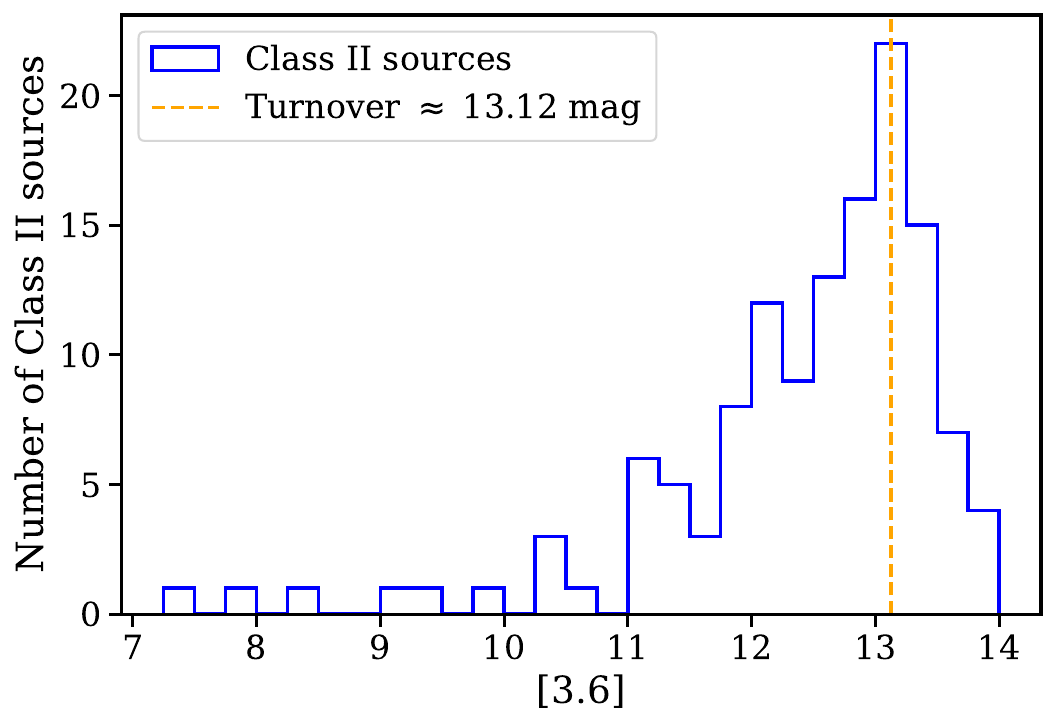} 
    \caption{The 3.6 $\mu$m magnitude histogram of the 130 Class II YSOs associated with G038.925-00.355. The orange dashed line marks the turnover magnitude (13.12 mag), which is adopted as the photometric completeness limit for this cloud.}
    \label{fig: Class II histogram}
\end{figure}

\subsection{Stellar Mass and Star Formation Rate} 

Once the YSOs associated with each cloud were identified, it was necessary to account for the low-mass population that remains undetected because of observational sensitivity limits. For this purpose, the 3.6 $\mu$m magnitude histogram of the Class II YSOs was constructed and the turnover (peak) magnitude was adopted as the photometric completeness magnitude \citep{Willis2013Completeness}. The 3.6 $\mu$m band was chosen because all identified YSOs in our sample were detected at this wavelength, providing a uniform and homogeneous basis for the completeness analysis across all clouds. Furthermore, the 3.6 $\mu$m band is the most sensitive of the IRAC bands and therefore offers the deepest and most complete census of the YSO population.

The corresponding mass completeness limit was determined using the PARSEC\footnote{\url{https://stev.oapd.inaf.it/PARSEC/tracks_v12s.html}} \citep{Bressan2012PARSEC} Pre-Main Sequence (PMS) isochrone for an age of 2 Myr, assuming a visual extinction ($A_V$) of 1 mag per kpc of the cloud distance \citep{Stahler2004Book}. The \cite{Kroupa2002IMF} initial mass function (IMF) defined as $\xi (M)=A\,M^{-2.3}$ for $M>0.5\,M_{\odot}$ and $\xi (M)=B\,M^{-1.3}$ for $M<0.5\,M_{\odot}$ was then normalized using the number of YSOs (Class I and Class II) detected above this completeness mass and extrapolated down to the lower-mass  limit ($0.08\,M_{\odot}$) to estimate the total YSO population.

The total stellar mass was calculated by multiplying the estimated total number of YSOs by a mean stellar mass of 0.41 $M_{\odot}$, \citep{Kirkpatrick2024_mean_mass}. Finally, the SFR was estimated by dividing the total stellar mass by the characteristic disk lifetime, assumed to be 3 Myr \citep{Pecaut&Mamajek2016disk_lifetime, Williams2011_disk_lifetime}.

As an illustration of the procedure we provide the details of the cloud G038.925-00.355. This particular cloud contains 47 Class I and 130 Class II YSOs. The 3.6 $\mu$m magnitude histogram of Class II sources is shown in Figure~\ref{fig: Class II histogram}. The turnover magnitude was found to be 13.12 mag corresponding to an absolute magnitude of 1.05 mag. A total of 125 Class I and Class II sources were brighter than this limit. The corresponding mass completeness limit derived from the PARSEC isochrone was 2.32 $\rm M_{\odot}$. Normalization and extrapolation of the Kroupa IMF yielded an estimated total YSO population of 3905 sources. This corresponds to a total stellar mass of $\sim1600$ M$_{\odot}$ and SFR of $0.53\times 10^{-3}$ M$_{\odot}$yr$^{-1}$.

\begin{deluxetable*}{lcccccccc}[htb!]
\tablecaption{Properties of our Sample Star Forming Clouds}
\tablewidth{0pt}
\tabletypesize{\footnotesize}
\tablehead{
\colhead{Cloud} & \colhead{Area} & \colhead{$Z$} & \colhead{$\rm N(H_2)$} & \colhead{M$_{\rm Cloud}$} & $\rm N_{\rm Class\,I}$ & $\rm N_{\rm Class\,II}$ & \colhead{SFR} & \colhead{SFE}\\
& \colhead{(pc$^2$)} &  & \colhead{($10^{21}$ cm$^{-2}$)} & \colhead{($10^{5}$ M$_\odot$)} & & & \colhead{($10^{-3}$ M$_\odot$ yr$^{-1}$)} & \colhead{($10^{-2}$)} 
}
\startdata
G010.560-00.080 & 26341.39 & 0.0392 & 3.54 & $20.94\pm3.74$ & 119 & 410 & $35.93\pm1.80$ & $4.90\pm0.86$ \\
G012.774+00.330 & 8853.09 & 0.0315 & 2.70 & $5.34\pm0.65$ & 25 & 106 & $30.40\pm3.34$ & $14.59\pm2.04$ \\
G012.870-00.255 & 2341.52 & 0.0391 & 3.54 & $1.85\pm0.18$ & 311 & 1229 & $7.13\pm0.23$ & $10.34\pm0.96$ \\
G013.272+00.066 & 2975.88 & 0.0440 & 1.66 & $1.11\pm0.14$ & 308 & 1011 & $14.44\pm0.55$ & $28.05\pm2.67$ \\
G013.913+00.239 & 381.29 & 0.0418 & 1.51 & $0.13\pm0.02$ & 10 & 44 & $0.83\pm0.15$ & $16.05\pm3.01$ \\
G014.331-00.640 & 249.20 & 0.0275 & 5.47 & $0.31\pm0.03$ & 230 & 1053 & $0.88\pm0.03$ & $7.98\pm0.78$ \\
G016.285-00.171 & 3269.08 & 0.0396 & 2.03 & $1.49\pm0.24$ & 94 & 594 & $3.07\pm0.15$ & $5.82\pm0.92$ \\
G017.638+00.157 & 5485.42 & 0.0323 & 2.71 & $3.34\pm0.57$ & 12 & 45 & $9.44\pm1.47$ & $7.83\pm1.67$ \\
G018.371-00.382 & 708.70 & 0.0370 & 3.78 & $0.60\pm0.07$ & 37 & 175 & $0.91\pm0.08$ & $4.32\pm0.61$ \\
G018.824-00.467 & 2409.12 & 0.0407 & 2.52 & $1.36\pm0.17$ & 42 & 367 & $5.43\pm0.33$ & $10.70\pm1.32$ \\
G019.480+00.155 & 5212.49 & 0.0302 & 3.86 & $4.51\pm0.82$ & 19 & 42 & $11.16\pm1.97$ & $6.91\pm1.63$ \\
G021.596-00.161 & 139.05 & 0.0473 & 0.95 & $0.03\pm0.01$ & 1 & 7 & $0.19\pm0.11$ & $15.86\pm7.91$ \\
G023.253-00.240 & 4527.17 & 0.0386 & 3.30 & $3.35\pm0.43$ & 149 & 1011 & $5.36\pm0.19$ & $4.59\pm0.59$ \\
G024.426+00.221 & 6248.27 & 0.0454 & 2.60 & $3.65\pm0.45$ & 95 & 409 & $11.84\pm0.65$ & $8.87\pm1.09$ \\
G025.796+00.239 & 541.56 & 0.0442 & 2.41 & $0.29\pm0.04$ & 6 & 31 & $0.80\pm0.17$ & $7.57\pm1.75$ \\
G026.272-00.062 & 899.79 & 0.0435 & 2.37 & $0.48\pm0.09$ & 26 & 65 & $1.78\pm0.25$ & $10.07\pm2.09$ \\
G026.597-00.024 & 18.06 & 0.0294 & 3.34 & $0.01\pm0.01$ & 3 & 17 & $0.05\pm0.02$ & $10.37\pm4.26$ \\
G027.000+00.190 & 1331.16 & 0.0423 & 1.69 & $0.50\pm0.10$ & 27 & 102 & $2.21\pm0.25$ & $11.62\pm2.39$ \\
G027.996-00.412 & 3135.70 & 0.0335 & 1.83 & $1.29\pm0.27$ & 8 & 61 & $7.46\pm1.19$ & $14.82\pm3.35$ \\
G028.652+00.072 & 1924.52 & 0.0428 & 1.75 & $0.76\pm0.13$ & 35 & 133 & $4.17\pm0.46$ & $14.20\pm2.54$ \\
G029.832-00.100 & 2953.56 & 0.0413 & 4.14 & $2.75\pm0.28$ & 47 & 310 & $6.35\pm0.39$ & $6.49\pm0.73$ \\
G030.000-00.335 & 614.99 & 0.0374 & 1.77 & $0.24\pm0.05$ & 13 & 136 & $1.25\pm0.13$ & $13.31\pm2.54$ \\
G031.200+00.119 & 1971.36 & 0.0410 & 3.61 & $1.58\pm0.18$ & 69 & 238 & $5.02\pm0.35$ & $8.69\pm1.05$ \\
G032.473+00.204 & 1099.62 & 0.0328 & 3.43 & $0.85\pm0.14$ & 4 & 33 & $1.99\pm0.41$ & $6.61\pm1.62$ \\
G034.404+00.227 & 125.07 & 0.0278 & 4.77 & $0.13\pm0.02$ & 50 & 246 & $0.39\pm0.03$ & $7.95\pm1.31$ \\
G034.758-00.525 & 229.10 & 0.0316 & 4.59 & $0.24\pm0.05$ & 28 & 119 & $0.44\pm0.05$ & $5.27\pm1.14$ \\
G035.063+00.330 & 6384.52 & 0.0337 & 3.48 & $4.98\pm1.03$ & 26 & 200 & $8.80\pm0.85$ & $5.03\pm1.09$ \\
G035.480+00.135 & 398.62 & 0.0368 & 2.27 & $0.20\pm0.04$ & 28 & 94 & $1.17\pm0.16$ & $14.78\pm3.08$ \\
G038.925-00.355 & 228.84 & 0.0308 & 2.32 & $0.12\pm0.01$ & 47 & 130 & $0.53\pm0.05$ & $11.84\pm1.57$ \\
G041.184-00.210 & 3209.91 & 0.0332 & 1.94 & $1.40\pm0.22$ & 21 & 96 & $3.52\pm0.38$ & $7.03\pm1.25$ \\
G041.740+00.095 & 703.05 & 0.0253 & 2.17 & $0.34\pm0.07$ & 5 & 19 & $1.33\pm0.33$ & $10.45\pm3.07$ \\
G043.149+00.028 & 1594.16 & 0.0246 & 6.45 & $2.31\pm0.22$ & 9 & 28 & $2.9\pm0.62$ & $3.63\pm0.82$ \\
G043.225-00.312 & 127.71 & 0.0304 & 1.38 & $0.04\pm0.01$ & 4 & 71 & $0.14\pm0.02$ & $9.54\pm2.59$ \\
G048.599+00.044 & 1137.79 & 0.0264 & 3.18 & $0.81\pm0.12$ & 11 & 32 & $1.32\pm0.33$ & $4.63\pm1.28$ \\
G049.428-00.464 & 2085.90 & 0.0310 & 6.28 & $2.94\pm0.29$ & 67 & 180 & $3.55\pm0.30$ & $3.50\pm0.44$ \\
G307.610-00.382 & 251.82 & 0.0289 & 2.21 & $0.12\pm0.02$ & 2 & 10 & $0.34\pm0.15$ & $7.55\pm3.37$ \\
G310.143+00.758 & 2437.59 & 0.0310 & 0.99 & $0.54\pm0.07$ & 32 & 260 & $3.60\pm0.30$ & $16.54\pm2.19$ \\
G312.598+00.048 & 3074.65 & 0.0319 & 1.58 & $1.09\pm0.20$ & 45 & 318 & $4.42\pm0.27$ & $10.86\pm1.90$ \\
G321.890-00.009 & 145.79 & 0.0296 & 2.05 & $0.07\pm0.01$ & 17 & 68 & $0.35\pm0.03$ & $13.50\pm2.25$ \\
G327.129-00.257 & 19279.27 & 0.0350 & 1.78 & $7.68\pm1.36$ & 88 & 557 & $29.90\pm1.58$ & $10.46\pm1.73$ \\
G328.221-00.531 & 791.53 & 0.0326 & 2.75 & $0.49\pm0.07$ & 76 & 569 & $2.07\pm0.11$ & $11.26\pm1.45$ \\
G331.123-00.530 & 1440.14 & 0.0275 & 2.75 & $0.89\pm0.13$ & 51 & 393 & $2.06\pm0.13$ & $6.50\pm0.96$ \\
G338.911+00.615 & 759.58 & 0.0396 & 2.33 & $0.40\pm0.06$ & 19 & 128 & $0.84\pm0.08$ & $5.95\pm1.03$ \\
G339.010-00.138 & 4856.12 & 0.0491 & 0.53 & $0.57\pm0.11$ & 45 & 362 & $9.34\pm0.55$ & $32.88\pm4.59$ \\
G343.873-00.078 & 45406.94 & 0.0330 & 2.26 & $22.91\pm4.28$ & 121 & 723 & $78.83\pm3.69$ & $9.36\pm1.63$ \\
\enddata
\tablecomments{Column (1): Cloud name. Column (2): Area. Column (3): Metallicity (Z).
Column (4): Mean column density. Column (5): Cloud mass with uncertainty. Columns (6)–(7): Number of Class I and Class II sources.
Column (8): Star Formation Rate (SFR) with uncertainty.
Column (9): Star Formation Efficiency (SFE) with uncertainty.}
\label{tab:CloudMass}
\end{deluxetable*}

\section{Analysis and Results} \label{sec: Results}

We now derive key parameters including SFR and SFE. We then investigate how these quantities vary with the physical properties of the molecular clouds.

\begin{figure}[t]
    \centering
    \includegraphics[width=1.0\columnwidth, height=0.3\textheight]{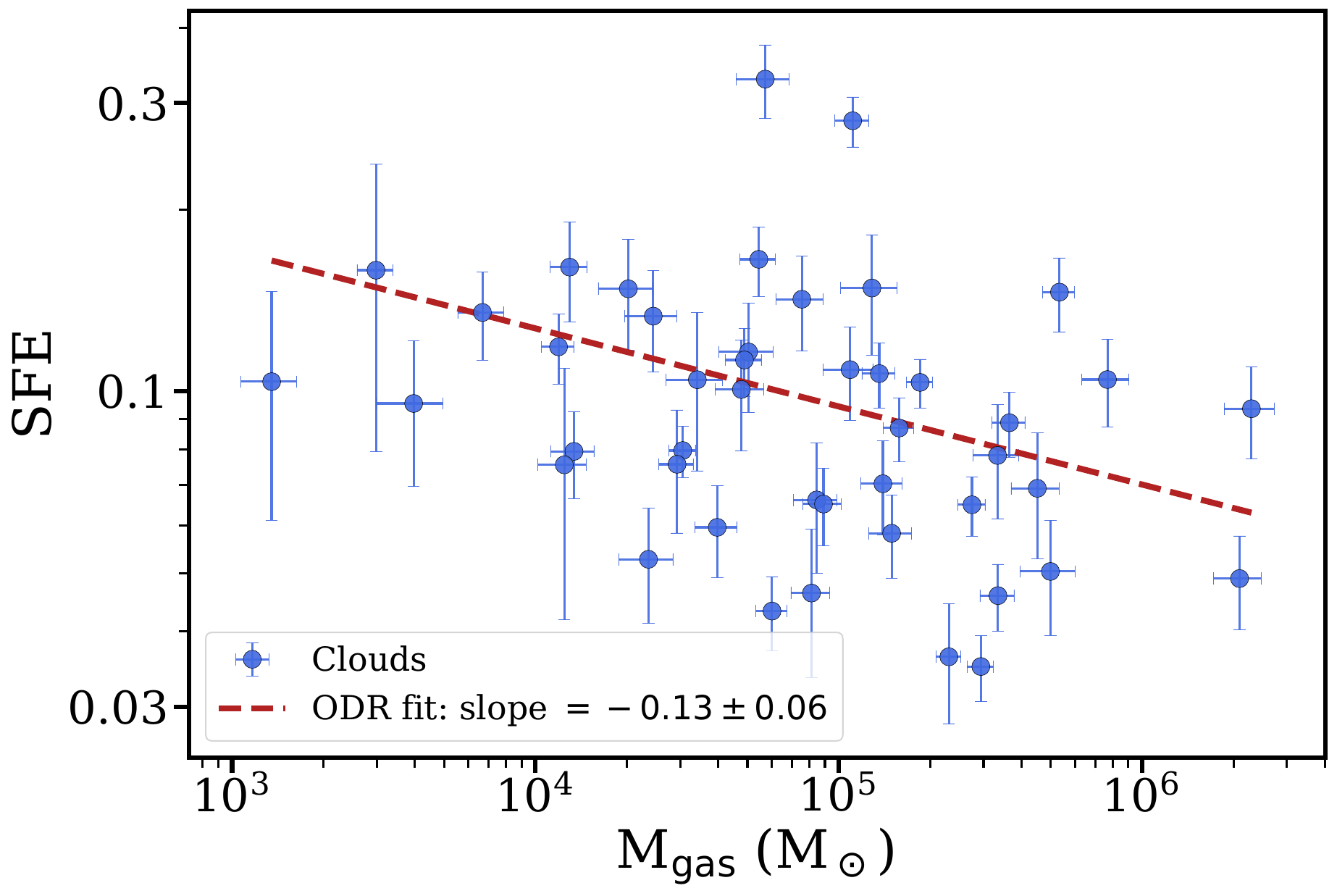} 
    \caption{Star formation efficiency as a function of gas mass for the molecular clouds in our sample. The blue circles represent individual clouds, while the red dashed line shows the ODR fit. See Table \ref{tab:fit_results} for details.}
    \label{fig: SFE_vs_Mgas_centred}
\end{figure}

\subsection{Variation of SFE with Cloud Mass}

The Star Formation Efficiency (SFE) in a star forming cloud is defined as the fraction of available gas mass that has been converted into stars \citep{Myers1986SFEdef,Shukirgaliyev2017SFEdef}

\begin{equation}
    \mathrm{SFE=\frac{M_*}{M_*+M_{gas}}}
\end{equation}

The SFE values for our clouds are listed in Table \ref{tab:CloudMass}. The derived SFE values span from $0.03-0.33$, with a median value of 0.09. These values are systematically higher than those reported in several previous studies on Galactic molecular clouds \citep{Evans2009LAW,Das2021NANComplex}. This difference may partly reflect the physical characteristics of our cloud sample. Variations in metallicity across the Galactic disk can influence the molecular gas properties and the efficiency of star formation. In addition, our sample was selected from molecular clouds associated with H{\sc ii} regions, which are sites of recent massive star formation. Such a selection preferentially targets actively star forming and relatively evolved clouds, which are expected to exhibit higher SFEs.

We fit the following relation to our data using orthogonal distance regression (ODR)

\begin{equation}
    y=N(x-x_0)+b,
    \label{eq: fit}
\end{equation}

where $x=\log(\rm M_{\rm cloud})$, $y=\log(\mathrm{SFE})$ and $x_0$ is the median value of $\log(\rm M_{\rm cloud})$. The fit shown in Figure~\ref{fig: SFE_vs_Mgas_centred} yields a slope of $N=-0.13\pm0.06$, indicating that the star formation efficiency decreases weakly with increasing cloud mass. The statistics (Pearson and Spearman correlation coefficients) are shown in Table \ref{tab:fit_results}, demonstrating a statistically significant negative trend despite the considerable scatter in the data. A plausible interpretation of this trend is that the gas directly participating in star formation does not increase in proportion to the total cloud mass. As a result, more massive clouds may contain increasingly large reservoirs of gas that are not actively forming stars, causing the global SFE to decrease with increasing cloud mass.

\begin{figure}[t]
    \centering
    \includegraphics[width=1.0\columnwidth, height=0.3\textheight]{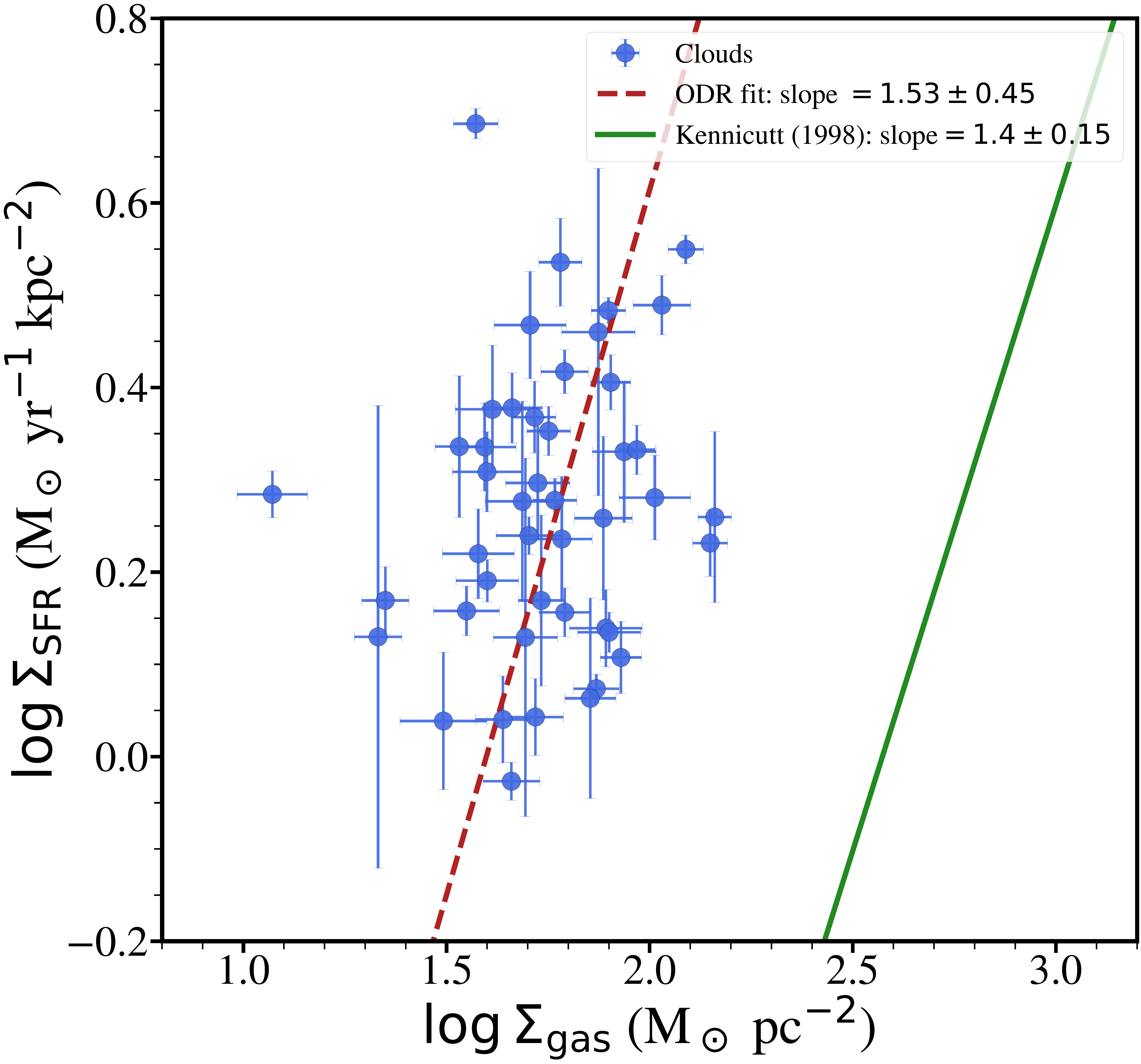} 
    \caption{Gas mass density and SFR density plot for the molecular clouds in our sample. The blue points represent individual clouds with their associated uncertainties. The red dashed line shows ODR fit to the data and the solid green line indicates the canonical \cite{Kennicutt1998LAW} relation. See Table \ref{tab:fit_results} for details.}
    \label{fig: KS plot}
\end{figure}

\subsection{Variation of SFR with Cloud Mass}

To investigate the star formation scaling relation in our cloud sample, we fitted a power-law relation between $\Sigma_{\rm gas}$ and $\Sigma_{\rm SFR}$, following Equation~\ref{eq: fit} with $x=\log(\Sigma_{\rm gas})$ and $y=\log(\Sigma_{\rm SFR})$. The fit yields a slope of $N=1.53\pm0.45$. However, the correlation coefficients (Table \ref{tab:fit_results}) do not indicate a statistically significant correlation, suggesting that the large uncertainty in the fitted slope reflects substantial cloud-to-cloud scatter rather than a robust underlying trend. Such scatter may arise from variations in cloud structure, dense-gas fraction, evolutionary stage, and local environmental conditions.

For reference, Figure~\ref{fig: KS plot} also shows the classical KS relation from \citet{Kennicutt1998LAW}, using the normalization and slope ($N=1.4 \pm 0.15$) reported in that work. However, this relation was derived from galaxy-averaged surface densities measured over kiloparsec scale regions \citep{Kennicutt1998LAW}, whereas our measurements correspond to resolved Galactic molecular clouds. Therefore, the comparison should not be interpreted as a direct one to one correspondence between the two datasets. Instead, the KS relation is included only as a reference to place our results in the broader context of star formation scaling relations. As seen in Figure~\ref{fig: KS plot}, our clouds generally exhibit higher $\Sigma_{\rm SFR}$ than predicted by the galaxy-averaged relation at a given $\Sigma_{\rm gas}$, consistent with previous studies of Galactic clouds and clumps. This offset is commonly attributed to the fact that resolved Galactic measurements preferentially trace actively star-forming molecular gas, whereas extragalactic measurements average over both star-forming and non-star-forming gas within large spatial regions. Subsequent extragalactic studies focusing primarily on molecular gas have reported a nearly linear relation ($N \approx 1$) in nearby galaxies \citep{Bigiel2008LAW,Leroy2013LAW}.

Several Galactic investigations have also explored analogous cloud and clump scale star formation relations. \cite{Wu2005LAW} studied 47 dense clumps traced by HCN emission and found an approximately linear relation between gas mass and star formation activity. Other studies \citep{Evans2009LAW,Heiderman2010LAW, Das2021NANComplex} investigated nearby molecular clouds using YSO based star formation rates. \cite{Rawat2025LAW} performed a similar study and reported a slope $N = 1.46 \pm 0.28$ for Galactic clumps.

\begin{figure}[t]
    \centering
    \includegraphics[width=1.0\columnwidth, height=0.3\textheight]{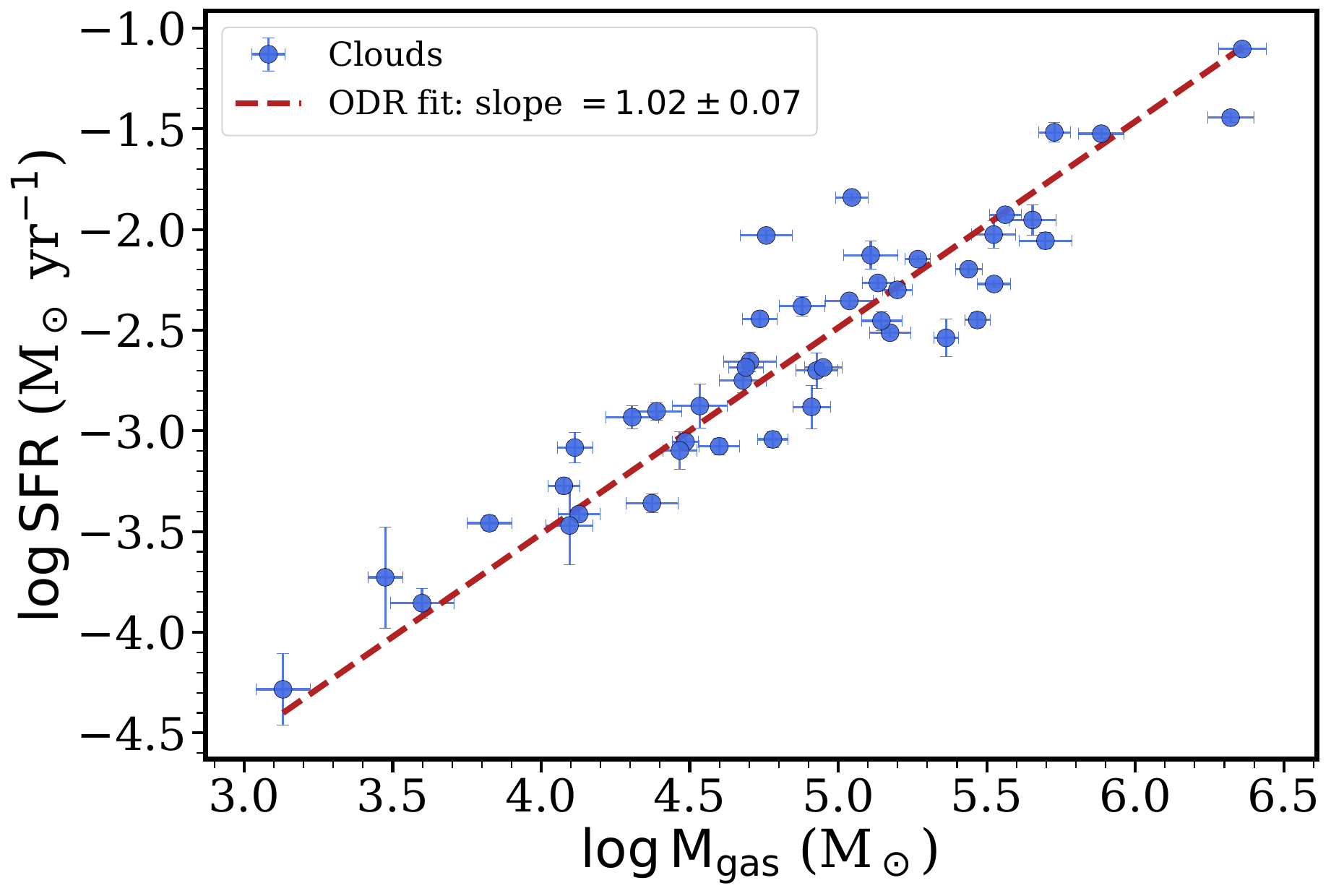} 
    \caption{SFR versus gas mass for the molecular cloud sample. The blue points represent individual clouds with their associated uncertainties. The red dashed line represents the best-fit ODR power-law relation. The near-unity slope indicates an approximately linear scaling between SFR and cloud mass over the mass range probed by our sample. See Table \ref{tab:fit_results} for details.}
    \label{fig: SFR_vs_Mgas}
\end{figure}

For our cloud sample, SFR ranges from $5.2\times10^{-5}$ to $7.9\times10^{-2}$ $\mathrm{M_{\odot} yr^{-1}}$, with a median value of $2.9\times10^{-3}$ $\mathrm{M_{\odot} yr^{-1}}$. The corresponding cloud masses span $10^3$ to $2.29\times 10^6$ $\mathrm{M_{\odot}}$. The power law relation between SFR and cloud mass was fitted using ODR following Equation~\ref{eq: fit}, with $x=\log(\rm M_{\rm gas})$ and $y=\log(\rm SFR)$. The best-fit slope of $N=1.02\pm0.07$ (Figure~\ref{fig: SFR_vs_Mgas}), indicating an approximately linear relation between cloud mass and star formation rate. This result suggests that the amount of star formation increases nearly in proportion to the available molecular gas mass over the mass range sampled in this work. The tight correlation and near unity slope are broadly consistent with the expectation that molecular gas serves as the primary reservoir for star formation. Similar sub linear to linear SFR gas mass relations have been reported in previous cloud-scale studies, for example, \citet{Vutisalchavakul2016SFR} found a slope of $N=0.66\pm0.12$, while \citet{Lada2010SFR} reported an approximately linear relation.

\begin{figure}[t]
    \centering
    \includegraphics[width=1.0\columnwidth, height=0.70\textheight]{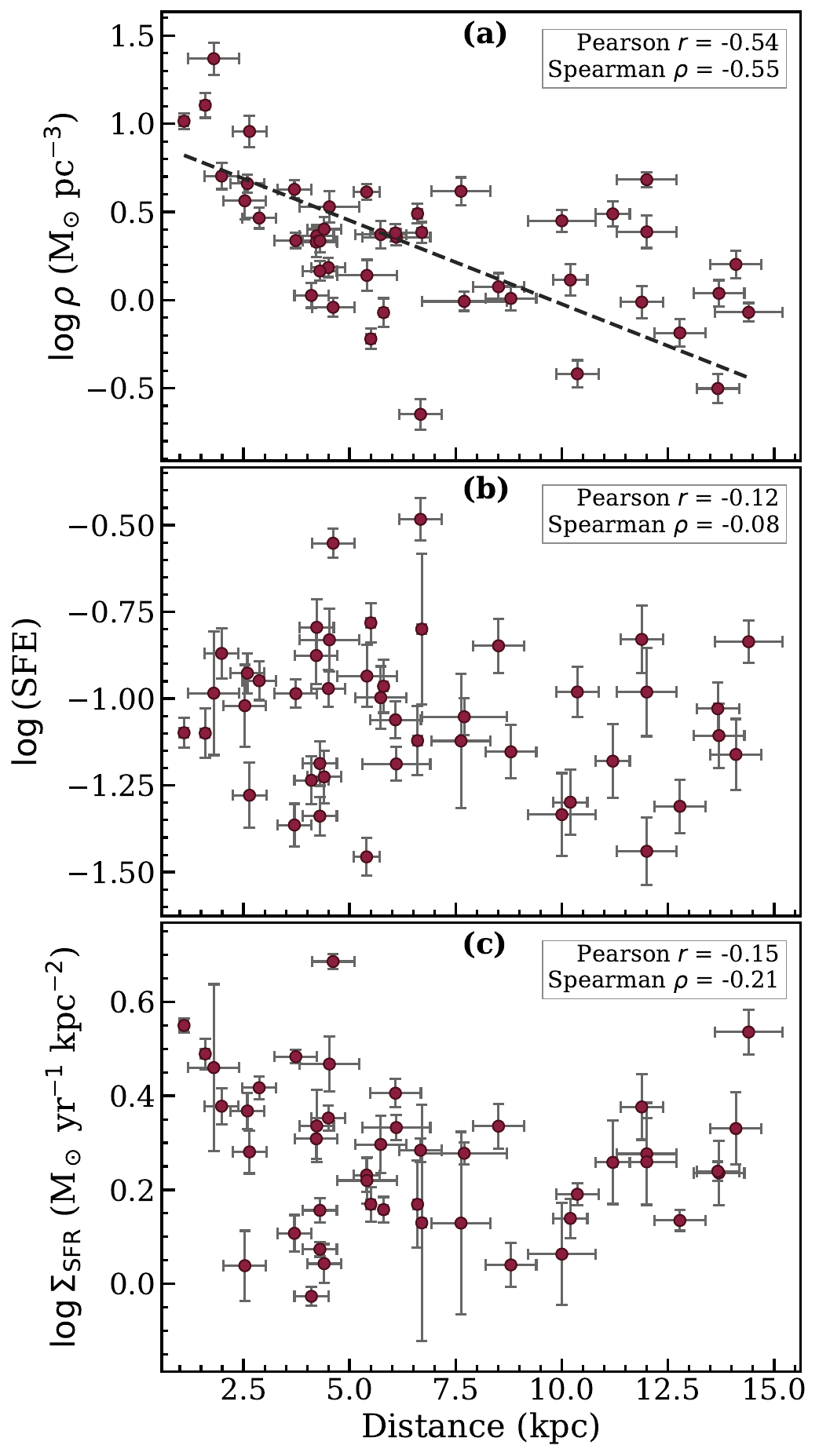} 
    \caption{Heliocentric distance ($d$) versus (a) mean density ($\rho$), (b) SFE and (c) $\Sigma_{\rm SFR}$ for the molecular clouds in our sample. The black dashed line in panel (a) shows the ODR fit. The plots are used to examine possible distance-dependent trends in the derived cloud properties.}
    \label{fig: Density vs density,SFE,Sigma_SFR}
\end{figure}

\subsection{The Volumetric Star Formation Law}

It is noted that the Kennicutt-Schmidt relation does not explicitly account for the three dimensional structure of star forming gas or the timescale of gravitational collapse. Motivated by theoretical models in which star formation is regulated by gravity, turbulence, and feedback, a volumetric formulation is adopted \citep[e.g.,][]{Krumholz2012Freefall}. This formulation relates the star-formation rate to the local gas surface density and the local free-fall time ($\mathrm{t_{ff}}$). In order to test the volumetric star formation law for our sample we calculate the free-fall time as $\mathrm{t_{ff}}=\sqrt{3\pi/32G\rho}$ with mean density $\rho=3\sqrt{\pi}\,\mathrm{M_{gas}}/4A^{3/2}$, assuming a spherical geometry \citep{Pokhrel2021LAW}.

Since our sample spans a wide range of heliocentric distances (1.1--14.4 kpc), we examined whether the derived mean densities are affected by distance-dependent resolution effects. Figure~\ref{fig: Density vs density,SFE,Sigma_SFR} shows that the mean density exhibits a moderate negative correlation (Table \ref{tab:fit_results}) with heliocentric distance, indicating that the derived densities decrease systematically with increasing distance. In contrast, neither the SFE nor the $\Sigma_{\rm SFR}$, which are used to examine the volumetric star formation law, shows a statistically significant correlation (Table \ref{tab:fit_results}) with heliocentric distance. Therefore, the volumetric star formation relations presented below should be interpreted with the understanding that the inferred free-fall times may be influenced by this distance dependence. However, the star formation observables themselves do not exhibit a corresponding dependence on heliocentric distance.

\begin{figure}[t]
    \centering
    \includegraphics[width=1.0\columnwidth, height=0.31\textheight]{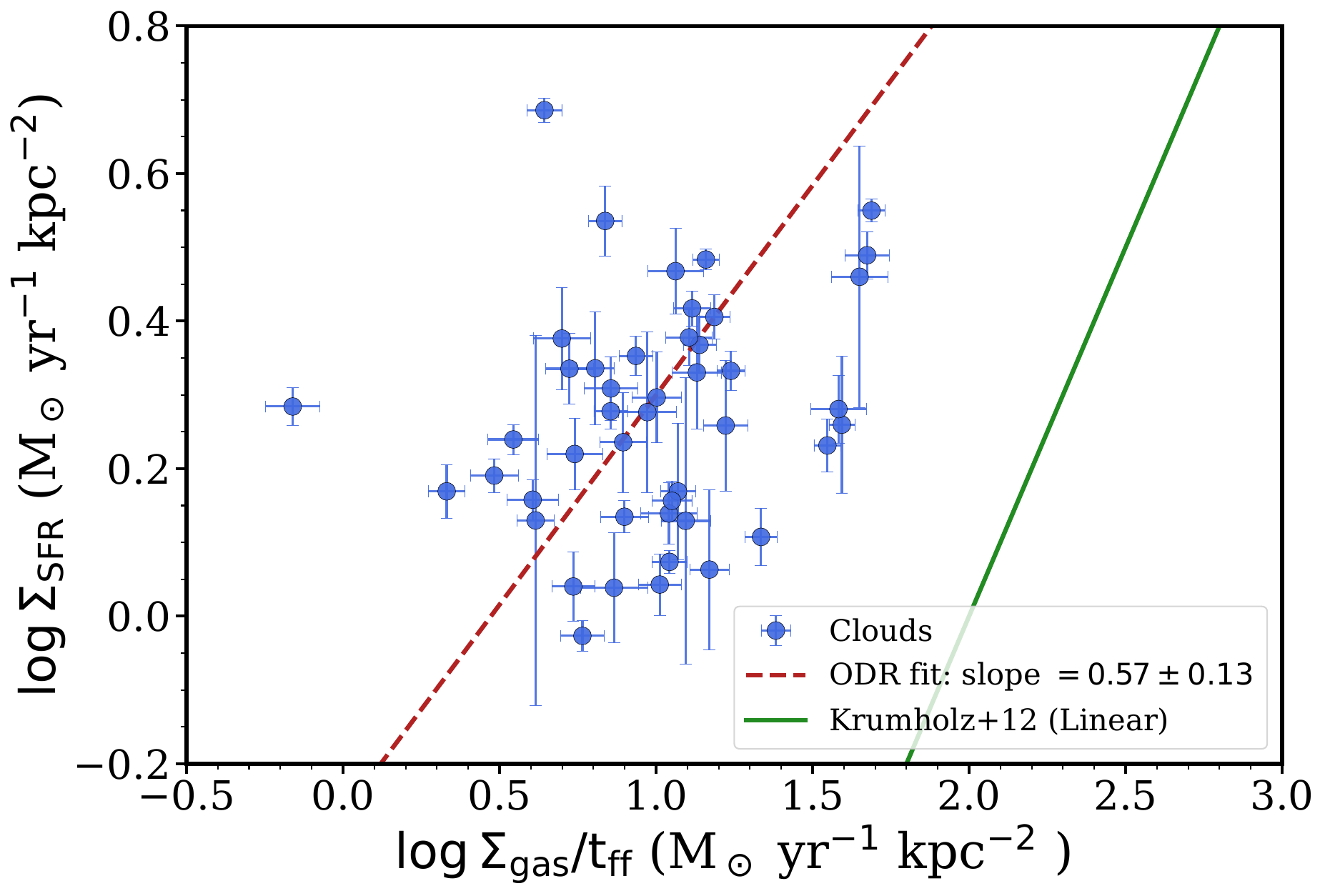} 
    \caption{Variation of $\mathrm{\Sigma_{SFR}}$ with $\mathrm{\Sigma_{gas}/t_{ff}}$ for the cloud sample. The red dashed line shows the ODR fit, while the solid green line denotes the linear volumetric star-formation law of \citet{Krumholz2012Freefall} with $\mathrm{\epsilon_{ff}}=0.01$. See Table \ref{tab:fit_results} for details.}
    \label{fig: Sigma_gas_vs_SigSFR_tff}
\end{figure}

Figure~\ref{fig: Sigma_gas_vs_SigSFR_tff} shows that the cloud sample appears systematically offset from the linear $\epsilon_{\rm ff}=0.01$ relation predicted by the volumetric star formation model of \citet{Krumholz2012Freefall}. The corresponding ODR fit, obtained using Equation~\ref{eq: fit}, gives a power-law slope of $N=0.57\pm0.13$. Nevertheless, the correlation statistics (Table \ref{tab:fit_results}) indicate that the observed relation is not statistically significant. For comparison, \citet{Pokhrel2021LAW} found a nearly linear relation for nearby molecular clouds, while \citet{Rawat2025LAW} reported a slope of $0.80\pm0.15$ for cluster-forming clumps. However, the apparent trend between $\Sigma_{\rm SFR}$ and $\Sigma_{\rm gas}/\rm t_{ff}$ indicates that the gravitational free-fall timescale still remains an important parameter governing star formation in molecular clouds.

\begin{figure}[t]
    \centering
    \includegraphics[width=1.0\columnwidth, height=0.31\textheight]{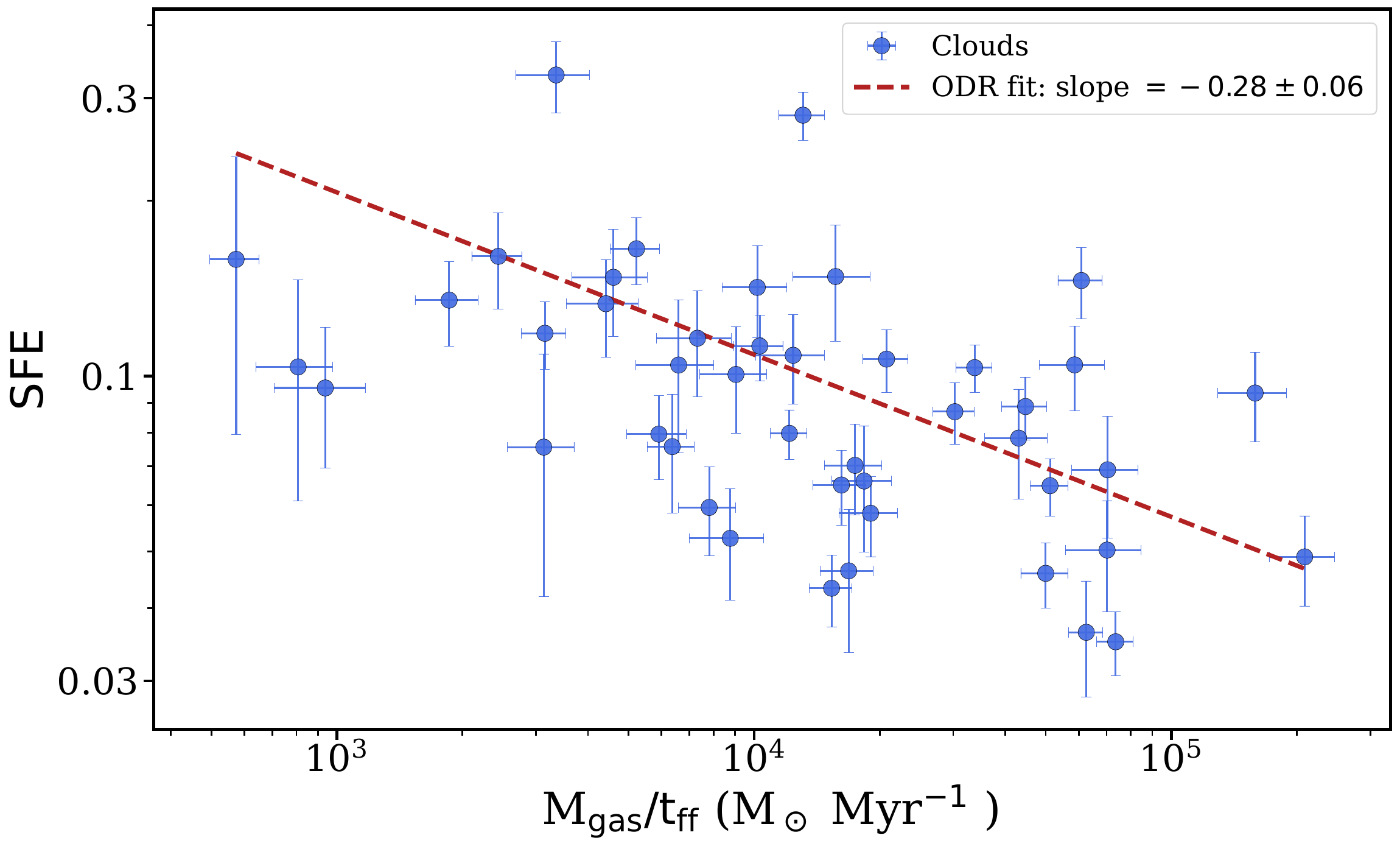} 
    \caption{SFE as a function of $\rm M_{gas}/t_{ff}$. The red dashed line shows the ODR fit, indicating a negative relationship between the two quantities. See Table \ref{tab:fit_results} for details.}
    \label{fig: SFE_vs_Mass_over_tff}
\end{figure}

To further investigate the role of the free-fall timescale, we examine the variation of SFE with $\rm M_{gas}/t_{ff}$, where $\rm M_{gas}/t_{ff}$ represents the amount of gas available for star formation per free-fall time. The ODR fit following Equation~\ref{eq: fit}, with $x=\log(\rm M_{gas}/t_{ff})$ and $y=\log(\rm SFE)$, gives a slope of $N=-0.28\pm0.06$ (Figure~\ref{fig: SFE_vs_Mass_over_tff}). Although the relation exhibits substantial scatter, a clear negative trend is present. This suggests that the fraction of gas converted into stars does not increase in proportion to the amount of gas available per free-fall time.

\begin{deluxetable*}{cccccc}
\tablecaption{Summary of the ODR fits and correlation analysis. Relevant relations were fitted in logarithmic space using $y=N(x-x_0)+b$, where $x_0$ is the median value of $x$. Pearson ($r$) and Spearman ($\rho$) correlation coefficients are listed with their corresponding $p$-values.
\label{tab:fit_results}}
\tabletypesize{\small}
\tablewidth{0pt}
\tablehead{
\multicolumn{2}{c}{Variables} & \colhead{Slope} & \colhead{Intercept} & \colhead{Pearson} & \colhead{Spearman} \\
\cline{1-2}
\colhead{$y$} & \colhead{$x$} & \colhead{$N$} & \colhead{$b$} & \colhead{$r$ ($p$)} & \colhead{$\rho$ ($p$)}}

\startdata
$\log(\mathrm{SFE})$ & $\log(\rm M_{\rm gas})$ & $-0.13\pm0.06$ & $-1.01\pm0.03$ & $-0.31$ (0.036) & $-0.35$ (0.017) \\
$\log(\Sigma_{\rm SFR})$ & $\log(\Sigma_{\rm gas})$ & $1.53\pm0.45$ & $0.21\pm0.06$ & $0.17$ (0.253) & $0.18$ (0.245) \\
$\log(\mathrm{SFR})$ & $\log(\rm M_{\rm gas})$ & $1.02\pm0.07$ & $-2.58\pm0.04$ & $0.94$ ($1.36\times10^{-21}$) & $0.91$ ($3.63\times10^{-18}$) \\
$\log(\rho)$ & $d$ & $-0.09\pm0.02$ & $0.38\pm0.05$ & $-0.54$ ($1.46\times10^{-4}$) & $-0.55$ ($8.64\times10^{-5}$) \\
$\log(\rm SFE)$ & $d$ & -- & -- & $-0.12$ ($0.417$) & $-0.08$ ($0.606$) \\
$\log(\Sigma_{\rm SFR})$ & $d$ & -- & -- & $-0.15$ ($0.314$) & $-0.21$ ($0.163$) \\
$\log(\Sigma_{\rm SFR})$ & $\log(\Sigma_{\rm gas}/\rm t_{\rm ff})$ & $0.57\pm0.13$ & $0.31\pm0.04$ & $0.23$ (0.134) & $0.24$ (0.112) \\
$\log(\mathrm{SFE})$ & $\log(\rm M_{\rm gas}/t_{\rm ff})$ & $-0.28\pm0.06$ & $-1.00\pm0.03$ & $-0.49$ ($6.86\times10^{-4}$) & $-0.53$ ($1.75\times10^{-4}$) \\
\enddata
\end{deluxetable*}

\section{Summary and Discussions} \label{sec: Discussions}

In this work, we present an analysis of star formation activity in the Milky Way molecular clouds using $^{12}$CO and $^{13}$CO spectral information from the FUGIN and ThrUMMS surveys. Gas masses are derived from the hydrogen column density maps, which are computed from the $^{13}$CO column density maps under the LTE assumption. YSOs are selected using the infrared color-color and color-magnitude criteria. The total stellar mass are estimated by using the 3.6 $\mu$m completeness limit, PARSEC tracks and a Kroupa IMF correction to account for the undetected low mass population.

Our sample consists of 45 clouds spanning the Galactic longitude $10^{\circ}<l<50^{\circ}$ and $300^{\circ}<l<350^{\circ}$. The heliocentric and the Galactocentric distances range from $1.1-14.4$ kpc and $3.1-8.2$ kpc respectively. We investigate the relationships between gas mass (and gas surface density) and SFR (and SFR density) for this sample. We now summarize and discuss the implications of our results. 

Our analysis shows that the clouds in our sample exhibit relatively high SFE values, reaching up to 0.33 in the most extreme case. While metallicity variations may contribute to the relatively high SFE values, it is also important to note that our sample is selected around H{\sc ii} regions. Consequently, the sample preferentially probes actively star forming and relatively evolved molecular clouds. Quiescent molecular clouds and earlier evolutionary stages such as infrared dark clouds (IRDCs) are therefore not represented in our sample. As a result, the star formation relations presented in this work should be interpreted as representative of molecular clouds associated with massive star formation rather than the complete Galactic molecular cloud population. The gas masses of our clouds span from $10^3$ M$_{\odot}$ to $22.91\times10^5$ M$_{\odot}$, and SFE shows a declining trend with increasing gas mass. This behavior suggests that the mass of dense gas directly involved in star formation does not increase monotonically with the total cloud mass.

A tight relation is observed in the SFR and cloud mass plane, with a power-law slope of $N=1.02\pm0.07$. This near-linear scaling implies that star formation activity closely tracks the amount of molecular gas available within a cloud, consistent with molecular gas being the primary fuel for star formation. By contrast, the surface-density relation yields a slope of $N=1.53\pm0.45$ and shows considerably greater scatter, consistent with the weak correlation statistics. It has been previously noted that in many cases, gas mass and star-formation in different regions of the Galaxy do not follow the star-formation law derived from local gas conditions (SFR varying linearly with the molecular gas surface density). In addition, the extra-galactic and high redshift star-formation exhibits complexities making it quite different from the near-universe star-formation law. It was thus proposed that star-formation might be a scale dependent phenomenon, governed by small scale physics, rather than some universal physical properties (large scale gravitational collapse) of the relevant astrophysical systems \citep{McKee2007Review}. Moreover, it is noted in the literature that the star-formation law can have biases that come from the specific methodology that is being adopted. For example, this may be related to data resolution, fitting methods, SFR and gas tracers, as well as YSO completeness \citep{Rawat2025LAW}.

Using numerical simulations as well as a large number of observations, \citet{Krumholz2012Freefall} show that SFR from Galactic to extra-galactic scales exhibit a consistent `local volumetric star-formation law' attributing the differences to be arising from varying geometry and three-dimensional sizes. They use a theoretical framework to show that SFR of a system is roughly 1\% of the molecular gas mass per local free-fall time \citep{Krumholz2012Freefall}. We note that the observed departure from the linear relation predicted by \citet{Krumholz2012Freefall} for $\epsilon_{\rm ff}=0.01$ suggest that our cloud sample does not strictly follow a universal efficiency per free-fall time relation.

\citet{k&M05} derive an analytic model of star-formation, leading to the Kennicutt-Schmidt law, by assuming supersonic turbulence in molecular clouds. Under the assumption of a lognormal density distribution (supersonic isothermal turbulence) and the notion that the density of the gas allows its gravitational potential energy to exceed the energy in turbulent motions, \citet{k&M05} argue that the turbulent velocity dispersion has a self-similar relation for a given length scale given by $l^{p}$, characteristic of supersonic turbulence. The sonic length of the system with respect to the Jeans length remains well-correlated with the SFR per free-fall time \citep[and references thereof]{k&M05}. Thus, universality in SFR scaling occurs when the free-fall time is considered in addition to local gas densities. However, we note that the derived values of $\mathrm{t_{ff}}$ and SFR depend on assumptions regarding cloud geometry, choice of gas tracers, YSO completeness etc. 

Recently, \citet{Zamoraetal24} performed a simulation of filamentary molecular clouds to test for the relative role of global gravitational contraction versus local turbulence in determining star-formation properties. Their study suggests that, lifetimes of clouds and their low star-formation efficiencies might be a direct consequence of global gravitational collapse, as opposed to the gravo-turbulence models of \citet{k&M05}. Using numerical simulations, \citet{Semadenietal24} propose that the long lifetime of clouds is driven by the constant replenishment of star-forming gas from low to high density regions in molecular clouds, until the onset of stellar feedback \citep{Dib2011SFE, Federrath2012simulation, Dib2013feedback, Dale2014HII_clouds, Geen2015feedback, Krumholz2019star_cluster, Suin2024feedback}. Furthermore, \citet{Zamoraetal24} argue that the lower values of the efficiency factor mainly come from two physical considerations, namely the method for determining the mass of young stars and the fact that the free-fall time of dense gas being a small fraction of the YSO lifetime (see \S 5 of \citealt{Zamoraetal24} for more details). 

It would hence be useful to lead observational studies for testing the proposed physical scenarios of star-formation in molecular clouds and the role of stellar feedback. \citet{Rawat2025LAW} studies the thermal free-free radio continuum emission of their clumps to investigate the role of massive stars in the regulation of SFR. They make an estimate of the radiation pressure to that of the gravitational pressure and show that the effect of stellar feedback is insignificant in their systems. While understanding of `star-formation' is a crucial question in astrophysics, it is a complex phenomenon involving gravity, fluid dynamics, turbulent gas motions, magnetic fields and feedback from stellar sources. It is possible that star-formation is a scale-dependent phenomenon, and its properties vary over different spatio-temporal scales. Recently \citet{krumholzetal25} suggested a new technique for measuring the differential virial co-efficient in molecular clouds to test for the gravo-turbulent model versus global gravitational collapse model. \citet{semadanietal26} extends the argument and use numerical simulations to investigate the intricacies, and scale dependencies of these physical situations. 

In this work, we have performed a comprehensive study of 45 clouds in the Milky-Way galaxy and have examined the local star-formation law for a larger range of cloud mass scales. It would thus be useful to utilize our sample of clouds (spanning a large mass range) to unravel the physical processes governing star-formation in the Universe.

\section*{Acknowledgements}
The authors sincerely thank the referee, Neal J. Evans, II, for his insightful comments and constructive suggestions, which significantly improved both the scientific content and the presentation of this paper. SC acknowledges financial support from ANRF through the POWER Fellowship (SPF/2022/000084). AS acknowledges financial support from UGC through UGC-NET Junior Research Fellowship. AH, TB and CM are thankful for the support of the S. N. Bose National Centre for Basic Sciences under the Department of Science and Technology, Government of India. AH also acknowledges the CSIR-HRDG, Government of India, for the funding of the fellowship. AS and SC thank Soumen Mondal, Sarita Vig, Saumyadip Samui, and Ritaban Chatterjee for useful discussions. AS and TB thank Sudeshna Patra for insightful discussions. SC also thanks the Inter University Center for Astronomy and Astrophysics for usage of their facilities through the IUCAA associateship program.

\bibliography{ref}{}
\bibliographystyle{aasjournal}

\end{document}